\documentclass[aps,pre,twocolumn,showpacs,amsmath,amssymb]{revtex4}
\makeatletter
\def\@dotsep{4.5}
\makeatother

\usepackage{graphicx}

\newlength{\onefig}
\newlength{\twofig}
\newcommand{\bq}{\begin{eqnarray}}
\newcommand{\eq}{\end{eqnarray}}
\newcommand{\bqn}{\begin{eqnarray*}}
\newcommand{\eqn}{\end{eqnarray*}}
\newcommand{\beq}{\begin{equation}}
\newcommand{\eeq}{\end{equation}}

\newcommand{\ve}[1]{\mathbf{#1}}

\begin{document}

\title{Density scaling in viscous liquids: From relaxation times to four-point susceptibilities}

\author{D. Coslovich}
\email{coslovich@cmt.tuwien.ac.at}
\affiliation{Institut f\"ur Theoretische Physik, Technische Universit\"at
  Wien, Wiedner Hauptstra{\ss}e 8-10, A-1040 Wien, Austria} 
\author{C. M. Roland}
\email{roland@nrl.navy.mil}
\affiliation{Naval Research Laboratory, Code 6120, Washington DC 20375-5342, USA} 
\thanks{Copyright (2009) American Institute of Physics. This article
  may be downloaded for personal use only. Any other use requires
  prior permission of the author and the American Institute of
  Physics.}

\date{\today}

\begin{abstract}
We present numerical calculations of a four-point dynamic susceptibility, 
$\chi_4(t)$, for the Kob-Andersen Lennard-Jones
mixture as a function of temperature $T$ and density $\rho$. 
Over a relevant range of $T$ and $\rho$, 
the full $t$-dependence of $\chi_4(t)$ and thus the maximum in $\chi_4(t)$,
which is proportional to the dynamic correlation volume, are invariant
for state points for which the scaling variable $\rho^{\gamma}/T$ is
constant. The value of the material constant $\gamma$ is the same as
that which superposes the relaxation time, $\tau$, of the system
versus $\rho^{\gamma}/T$. Thus, the dynamic correlation volume is
a unique function of $\tau$ for any thermodynamic condition in the
regime where density scaling holds. Finally, we examine the conditions
under which the density scaling properties are related to the
existence of strong correlations between pressure and energy
fluctuations.
\end{abstract}


\maketitle

During the last decade extensive evidence has accumulated that the
dynamics of molecules in supercooled, highly viscous liquids is
inherently heterogeneous
\cite{berthier_spontaneous_2007,ediger__2000,glotzer_spatially_2000};
that is, spatial variations in mobility persist for times commensurate
with the structural relaxation time $\tau$. Thus, there is
growing interest in characterizing the dynamic heterogeneities to
better understand the phenomena associated with the many-body dynamics
of vitryifing liquids. These phenomena include
rotational-translational decoupling
\cite{fujara_translational_1992,cicerone_do_1995}, the
dynamic cross-over
\cite{stickel_dynamics_1995,casalini_connection_2003}, enhanced
mobility under confinement
\cite{mckenna_size_2000}, non-exponentiality
of the relaxation~\cite{bohmer_nonexponential_1993}, and most
prominently, the slowing down of the translational and rotational
dynamics upon approach to the glass transition temperature $T_{g}$. To
explain these features, theories of the glass transition as diverse as
the classical Adam-Gibbs entropy~\cite{adam_temperature_1965} and
Cohen-Grest free volume~\cite{cohen_dispersion_1981} models, as well
as more modern approaches~\cite{lubchenko_theory_2007,avramov_viscosity_2005},
invoke dynamic heterogeneity having a length scale that grows in
concert with $\tau$ upon drawing near to $T_{g}$.

A proper description of the dynamic heterogeneities requires 
multi-point dynamic susceptibilites, which reflect 
correlations in the spatial variation of the dynamics. A four-point 
dynamic susceptibility $\chi_4(t)$ can be calculated as the variance of the 
self-intermediate scattering function $F_s(k,t)$
\begin{equation}\label{eqn:chi4}
\chi_4(t) = N \left[ \left\langle f_s^2(k,t) \right\rangle 
  - F_s^2(k,t) \right]
\end{equation}
where $f_s(k,t)$ is the instantaneous value, such that $\langle
f_s(k,t)\rangle=F_s(k,t)$. $\chi_4(t)$ quantifies the amplitude of the
fluctuations associated with $f_s(k,t)$ and has a maximum,
$\chi_4^{max}=\chi_4(t_{max})$, proportional to the dynamic correlation
volume~\cite{donati_theory_2002,toninelli_what_2005}.  $\chi_4(t)$
exhibits various regimes~\cite{toninelli_what_2005}, but most
interesting for study of the glass transition is the behavior around
$t_{max} \sim \tau$~\cite{berthier_direct_2005}. Recent numerical and
simulation works have shown that the dynamic correlation volume grows
upon cooling
\cite{dalle-ferrier_spatial_2007,capaccioli_dynamically_2008,fragiadakis_density_2009}.
However, very little is known about the combined temperature and
\textit{density} dependences of $\chi_4(t)$. In particular, unexplored
is the possibility of a description of the $\rho$ and $T$ dependences
of $\chi_4(t)$ in terms of the scaling property established for the
structural relaxation time
\cite{casalini__2004,alba-simionesco__2004,dreyfus__2004}
\begin{equation}\label{eqn:scaling}
\tau ={\mathcal F}_1(\rho^\gamma/T)
\end{equation}
where ${\mathcal F}_1$ is a function and $\gamma$ a material
constant. Experiments have shown that Eq.~\eqref{eqn:scaling}, with similar
relations for the diffusion constant and viscosity, applies universally
to organic, non-associated liquids~\cite{roland__2005}, with a range
of validity extending from the high temperature Arrhenius regime down
to $T_{g}$~\cite{casalini_liquids_2005}. We anticipate a
similar relation for $t_{max}$
and examine the possibility that this scaling property extends to the 
dynamic correlation volume
\begin{equation}
\chi_4^{max} ={\mathcal F}_2(\rho^\gamma/T)
\end{equation}

To cast this work in more general terms, we assess
whether for a prototypical model glass-former, the density scaling
properties hold for the full time dependence of both $F_s(k,t)$ and
$\chi_4(t)$, using the same scaling exponent $\gamma$.  The connection
between the dynamic properties of viscous liquids and $\gamma$ is
intriguing because molecular dynamics (MD) simulations have shown this
parameter to be a measure of the correlation between fluctuations in
the potential energy and the virial
\cite{pedersen_strong_2008,bailey_pressure-energy_2008-1,coslovich_pressure-energy_2009},
possibly reflecting a hidden scale invariance in viscous
liquids~\cite{schrder_pressure-energy_2009-1}. $\gamma$ is also
related to the steepness of the effective repulsive potential in the
range of closest-approach between
particles~\cite{roland__2006,coslovich_thermodynamic_2008,hall_intermolecular_2008},
although this connection may seem tenuous given the limitations of a
two-body potential in describing interactions in real liquids and the
possible non-trivial role of attractive
forces~\cite{hall_intermolecular_2008,bailey_pressure-energy_2008,berthier_nonperturbative_2009}.
We further analyze this herein by evaluating pressure-energy
correlations for the model studied
in~\cite{berthier_nonperturbative_2009}, which did not conform to
density scaling over the relevant density regime.

In this work we consider the well-studied Kob-Andersen (KA)
binary mixture~\cite{kob_testing_1995} as a model glass-forming liquid. It
consists of 1000 particles in a cubic box with periodic boundary
conditions. Particles interact through the Lennard-Jones (LJ)
potential
\begin{equation}\label{eqn:lj} 
u_{\alpha\beta}(r) = 4\epsilon_{\alpha\beta}\left[
  {\left( \frac{\sigma_{\alpha\beta}}{r} \right)}^{12} - {\left(
    \frac{\sigma_{\alpha\beta}}{r} \right)}^6 \right]
\end{equation}
where $\alpha, \beta = 1,2$ are species indices. The values of the
parameters in Eq.~\eqref{eqn:lj} can be found in the original paper
\cite{kob_testing_1995}. In the following we use reduced LJ units,
assuming $\sigma_{11}$, $\epsilon_{11}$, and
$\sqrt{m_1\sigma_{11}^2/\epsilon_{11}}$ (where $m_{1}$ is mass) as
units of distance, energy, and time, respectively. We performed
MD simulations in the NVT ensemble using the
Nos\'e-Poincar\'e thermostat~\cite{nose__2001} with a mass parameter
$Q$=5.0. We considered five isochoric paths in the density range 1.150
$\le $ $\rho$ $\le $ 1.350. For each state point we averaged
the dynamic properties over 20 independent realizations of the system.

\begin{figure}[tb]
\begin{center}
\includegraphics*[width=\onefig]{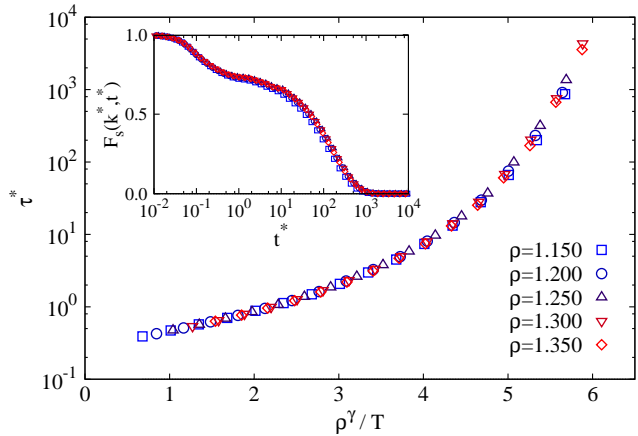}
\end{center}
\caption{\label{fig:fkt} Reduced relaxation times $\tau^*=\tau(\rho^{1/3}
  T^{1/2})$ as a function of $\rho^\gamma/T$ with $\gamma=5.1$ for
  all studied densities. Inset: self intermediate scattering functions as a
  function of reduced time $t^*=t(\rho^{1/3} T^{1/2})$ for state
  points at which $\rho^\gamma/T=5.07$: $T=0.402$ (at $\rho=1.15$),
  $T=0.50$ (at $\rho=1.20$), $T=616$ (at $\rho=1.30$), and $T=0.912$
  (at $\rho=1.350$). A constant reduced wave-vector
  $k^*=k(\rho^{1/3})=7.44$ is considered.}
\end{figure}

Following previous work and 
to emphasize the connection with the exact density scaling relations 
observed for inverse power law (IPL) potentials, we analyze 
``reduced'' quantities (indicated by stars) using $\rho$$^{-1/3}$ and $T^{1/2}$ as 
reduction parameters for distances and velocities, respectively. We begin our 
investigation of the density scaling properties of the KA model by 
calculating
$
F_s(k,t) =
(1/N)\sum_{j=1}^{N} \langle \exp
\left\{i \ve{k} \cdot [\ve{r}_j(t)-\ve{r}_j(0)]\right\} \rangle
$
for the various densities at a fixed
reduced wave-vector $k^*=k (\rho^{1/3})=7.44$, which
matches the position of the first peak in the static structure factor ($k=7.0$)
for the well-studied density $\rho=1.2$. The reduced
relaxation times $\tau^{*}$, defined as the time for
$F_{s}(k^{* },t^{* })$ to decay by a factor of $e$, are shown in
Fig.~\ref{fig:fkt} versus the scaling variable $\rho^{\gamma}/T$. 
The material constant $\gamma = 5.1 \pm 0.1$
provides the optimal collapse of $\tau^{*}$ onto a
single curve; this value is in accord with the scaling behavior found
previously for the diffusion coefficient of this system
\cite{coslovich_thermodynamic_2008}. We note that the
pressures attained at the lowest temperatures range from 1 to 20
reduced LJ units, depending on $\rho$. Assuming Argon units,
this corresponds to 0.04 to 0.8 GPa, which is a significant and experimentally
accessible pressure range.

\begin{figure}[tb]
\begin{center}
\includegraphics*[width=\onefig]{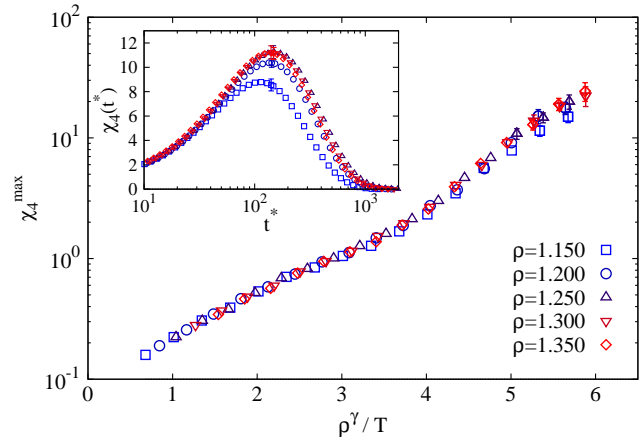}
\end{center}
\caption{\label{fig:chi4} Maximum of four point dynamic
  susceptibility as a function of $\rho^\gamma/T$ with
  $\gamma=5.1$ for all studied densities. Inset: four point
  dynamic susceptibility as a function of reduced time
  $t^*$ for state points at which $\rho^\gamma/T=5.07$ (same state
  points as inset of Fig.~\ref{fig:fkt}).}
\end{figure}

It has been demonstrated experimentally that the shape of the
frequency-dependent linear response function is a function of $\tau$
and thus of
$\rho^{\gamma}/T$~\cite{roland_isochronal_2003,ngai_do_2005}, a result
consistent with the existence of ``isomorphic'' points in liquid state
diagrams~\cite{gnan_pressure-energy_2009}. We calculated the
intermediate scattering functions for each of the five densities at
the respective temperatures corresponding to a fixed value of
$\rho^\gamma / T$ = 5.07. In the inset of Fig.~\ref{fig:fkt} these correlation
functions are plotted as a function of reduced time, and, as found
previously over a more limited density range~\cite{gnan_pressure-energy_2009}, $F_s(k^*,t^*)$ has essentially the
same shape for state points for which $\rho^{\gamma}/T$ is constant. Thus, not only do the
relaxation times superpose as a function of $\rho^{\gamma}/T$, but
the entire $t$-dependence of the correlation functions is invariant
for isomorphic state points.

We now examine the correlations in the dynamic fluctuations to assess
explicitly whether density scaling can be extended to high-order
correlation functions, as envisaged
in~\cite{gnan_pressure-energy_2009}. To do this
we calculate the four point dynamic susceptibility
[Eq.~\eqref{eqn:chi4}] associated with the complex instantaneous value
of $f_s(k,t)$ at the same fixed reduced wave-vector $k^{*}$
considered above. The use of the complex self intermediate scattering
function (rather than the real part used in previous studies) removes
the finite, long-time limit from $\chi_4(t)$ without altering the
general features of the correlation function. In Fig.~\ref{fig:chi4}
the maximum of $\chi_4(t)$ is plotted as a function of $\rho^\gamma/T$
using the aforementioned $\gamma =5.1$. To provide an estimate of the
statistical uncertainties, we include errors bars that represent two
standard deviations on the average over system realizations for
selected states. As seen in Fig.~\ref{fig:chi4}, density scaling
applies to $\chi_4^{max}$ within the estimated error, using the same
value for the scaling exponent that superposes the relaxation
times. In the inset of Fig.~\ref{fig:chi4}, we show the full
$t$-dependence of $\chi_4(t)$ for state points at which $\rho^\gamma/T=5.07$.
As is the case for the ``average''
intermediate scattering functions, within the estimated error bars the
$\chi_4(t)$ fall on a single curve. Deviations from the 
scaling are observed only for the lowest investigated density
($\rho=1.15$).

\begin{figure}[tb]
\begin{center}
\includegraphics*[width=\onefig]{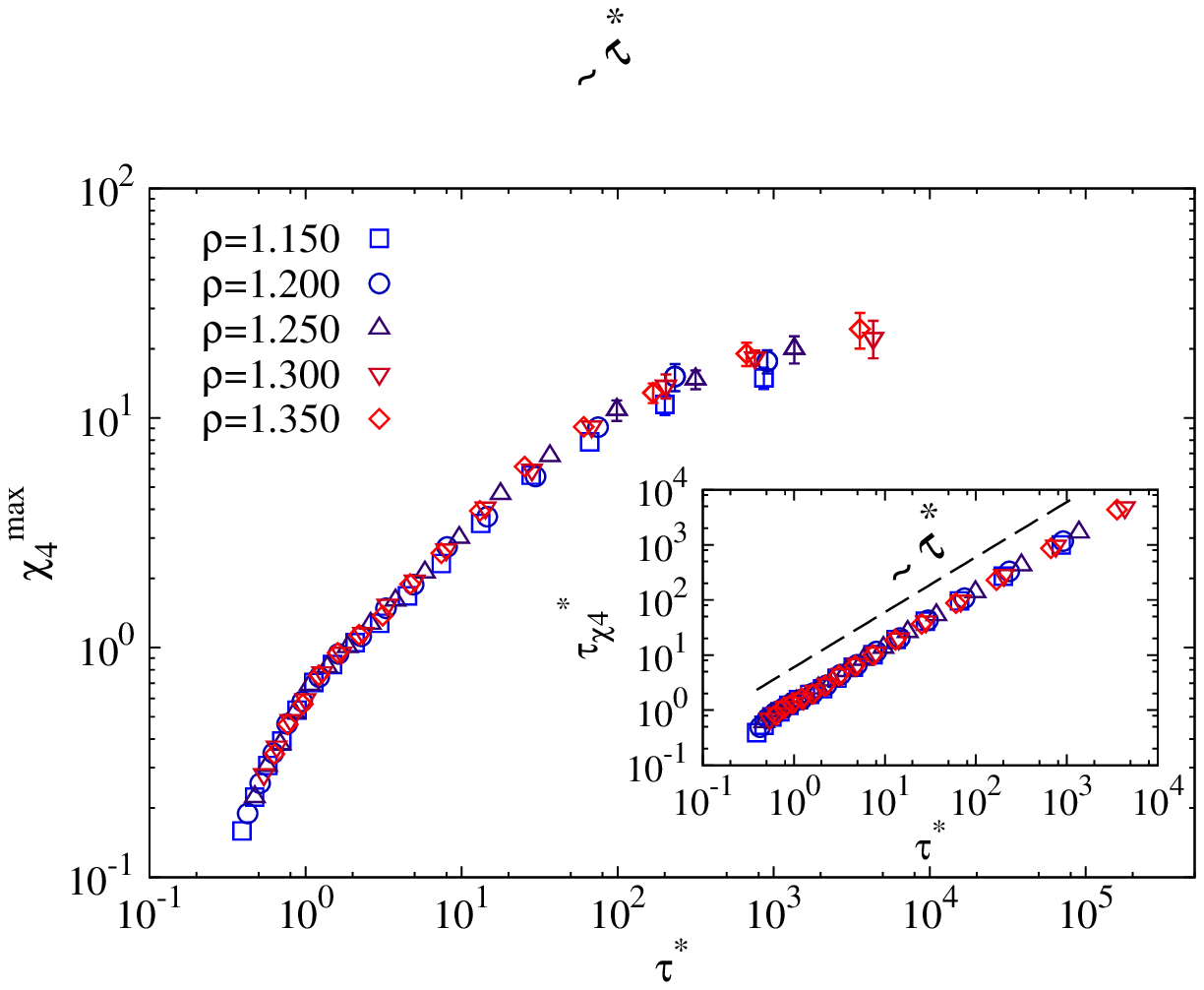}
\end{center}
\caption{\label{fig:tau_chi4} Maximum of four point dynamic
  susceptibility as a function of reduced relaxation
  times $\tau^*$ for all studied densities $\rho$. Inset: reduced time
  at which $\chi_4(t^*)$ is maximum as a function of $\tau^*$.}
\end{figure}

To make contact with previous numerical and experimental work on four
point dynamic susceptibilities, we show in Fig.~\ref{fig:tau_chi4}
$\chi_4^{max}$ as a function of the reduced relaxation times,
$\tau^{*}$. Since $\chi_4^{max}$ is proportional to the dynamic
correlation volume, Fig.~\ref{fig:tau_chi4} confirms the presence of a
steady but rather mild growth of dynamic correlations as the
structural relaxation times increase to the point of
vitrification. It also shows that, at fixed $\tau^{*}$, the dynamic
correlations are invariant to either $T$ or $\rho$, as expected from
the density scaling of both the average dynamics and the dynamic
correlations for the same value of $\gamma$. Recent
experiments~\cite{fragiadakis_density_2009}
have shown that a similar result holds for the temperature derivatives
of the two point dynamic correlation function, $\chi_{T}(t)$, which
provides a lower bound to
$\chi_4(t)$~\cite{dalle-ferrier_spatial_2007,berthier_direct_2005}. This
correspondence supports the validity of the experimentally accessible
$\chi_{T}(t)$ as an approximation to $\chi_{4}(t)$. In the inset of
Fig.~\ref{fig:tau_chi4}, we show the $\tau^*$-dependence of the
reduced time associated with the maximum in $\chi_4(t)$. As expected,
the two quantities are essentially equal. 

Recent numerical
work~\cite{pedersen_strong_2008,bailey_pressure-energy_2008-1,coslovich_pressure-energy_2009}
has shown that the dynamic scaling exponent $\gamma$ can be
independently estimated on the basis on the correlation between
fluctuations of two thermodynamic quantities, the potential energy,
$U$, and the virial, $W$ (the configurational part of the
pressure). These fluctuations, $\Delta U=U-\langle U\rangle$ and
$\Delta W=W-\langle W\rangle$ are proportional for particles
interacting with IPL potentials~\cite{pedersen_strong_2008} and have
been shown to be strongly correlated (Pearson correlation coefficients
$R>0.9$) for various other
liquids~\cite{bailey_pressure-energy_2008-1}. In the latter cases the
slopes, obtained from linear regression of $\Delta W$ vs. $\Delta U$,
are equal within the statistical fluctuations to the dynamic scaling
exponent~\cite{coslovich_pressure-energy_2009}. Such results support
the conjecture that liquids display strong $U$-$W$ correlations if and
only if they comply with density scaling~\cite{pedersen_strong_2008},
the inference being that these properties have a common origin in the
same generalized IPL approximation of the interaction
potential~\cite{bailey_pressure-energy_2008-1}.

\begin{figure}[tb]
\begin{center}
\includegraphics*[width=\onefig]{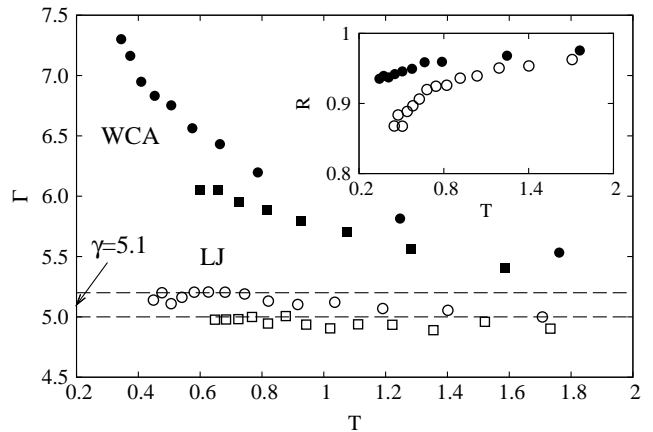}
\end{center}
\caption{\label{fig:uw} Slope of $U$-$W$ correlations as function of
  $T$ at $\rho=1.2$ (circles) and $\rho=1.3$ (squares) for the
  KA LJ model (open symbols) and its WCA variant (filled
  symbols). The dashed lines indicate the range of the dynamic scaling
  exponent $\gamma=5.1\pm 0.1$. Inset: Pearson correlation coefficient
  as a function of $T$ for $\rho=1.2$.}
\end{figure}

To further elaborate on these aspects, we show in Fig.~\ref{fig:uw}
the $\Delta W$ vs. $\Delta U$ slopes, $\Gamma$, obtained for the KA model, as
a function of $T$ for two densities ($\rho=1.2$ and
1.3). $\Gamma(T,\rho)$ is always close to the scaling exponent (within
the estimated uncertainty), as found previously from simulations along
isobaric paths~\cite{coslovich_pressure-energy_2009}. Also included in
Fig.~\ref{fig:uw} are results from additional simulations carried
out using the purely repulsive, Weeks-Chandler-Andersen (WCA) variant of the KA
model~\cite{weeks__1971,chandler_lengthscale_2006}. In the WCA model, the interaction parameters
are unchanged but each of the pair potentials, $u_{\alpha \beta}(r)$,
is shifted so that the minimum is zero and the potential is truncated at this
minimum~\cite{weeks__1971}. Very recently, Berthier and
Tarjus~\cite{berthier_nonperturbative_2009} concluded from simulations
of this model that density scaling surprisingly \textit{requires} the
contribution of the attractive interactions; in the WCA model, in
fact, the scaling of $\tau$ was absent except at very high
densities. As seen from Fig.~\ref{fig:uw} and its inset, in the WCA
model there are strong $U$-$W$ correlations ($R>0.9$); however,
the value of $\Gamma$ changes significantly as the state parameters
are varied. That is, the local scaling exponents change too much (for
reasons yet unknown), causing a breakdown of the density
scaling. Nevertheless, for every state point $U$ and $W$ are strongly
correlated.

In the light of these results, the conjecture of Pedersen et
al.~\cite{pedersen_strong_2008} connecting strong $U$-$W$ correlations
and density scaling must be partly reformulated. The existence of
strong $U$-$W$ correlations is not a sufficient condition for
Eq.~\eqref{eqn:scaling} to apply. There is an additional requirement,
that the slopes of the $U$-$W$ correlation must be (almost)
insensitive to variations of the state parameters. This is consistent
with previous simulation and experimental results on local dynamic
scaling exponents~\cite{le_grand_scaling_2007,win_glass_2006}. Under
these conditions, density scaling applies to a very good approximation to all time-dependent
properties, including high-order time-dependent correlation functions,
as conjectured in~\cite{gnan_pressure-energy_2009}.
On the other hand, further work is needed to understand under which
conditions the slopes of $U$-$W$ correlation are insensitive to
variations of $\rho$ and $T$, and how this is related to the
attractive part of the interaction potential.


From simulations over a range of $T$ and $\rho$, the
dynamic correlations in a supercooled LJ mixture are shown
to have a spatial extent that depends only on the quantity
$\rho^{\gamma}/T$. 
In consideration of the general behavior of
liquids obeying density scaling, this means the dynamic correlation volume is related to 
other mutually correlated properties: the relaxation time, the shape
of the relaxation function, $U$-$W$ correlations, and at least
approximately the isobaric fragility. From experiments it is also
found that the relaxation time is constant both at the onset of
non-Arrhenius behavior at high $T$~\cite{roland_characteristic_2008}
and at the dynamic crossover at \textit{ca.}
1.2$T_{g}$~\cite{casalini_dynamic_2003,casalini_viscosity_2004}. Since $\chi_4^{max}$ depends only on $\tau$,
these changes in the dynamics also occur at a
fixed (pressure-independent) correlation volume. 
Although examining repulsive exponents other than the LJ value of 12
remains for future work, if the equivalence of the scaling exponents for
$\chi_4^{max}$ and $\tau$ is maintained, we expect that the dynamic
correlation volume should vary among different materials.

Acknowledgements---D.~C. acknowledges financial support by the Austrian Science Fund
(FWF) (Project number: P19890-N16). The work at NRL was supported by
the Office of Naval Research.





\end{document}